# Two-Dimensional Rare Earth - Gold Intermetallic Compounds on Au(111) by Surface Alloying

Yande Que,[1*] Yuan Zhuang,[1] Ziyuan Liu,[2] Chaoqiang, Xu,[1] Bin Liu,[1] Kedong Wang,[3] Shixuan Du,[2] and Xudong Xiao[1*]

[1] *Department of Physics, the Chinese University of Hong Kong, Shatin, Hong Kong, China*

[2] *Institute of Physics and University of Chinese Academy of Sciences, Chinese Academy of Sciences, Beijing, 100190, China*

[2] *Department of Physics, Southern University of Science and Technology, Shenzhen, Guangdong 518055, China*

[*] Corresponding author.

Email: ydque@phy.cuhk.edu.hk and xdxiao@phy.cuhk.edu.hk

**Abstract:**

Surface alloying is a straightforward route to control and modify the structure and electronic properties of surfaces. Here, We present a systematical study on the structural and electronic properties of three novel rare earth-based intermetallic compounds, namely $ReAu_2$ (Re = Tb, Ho, and Er), on Au(111) via directly depositing rare-earth metals onto the hot Au(111) surface. Scanning tunneling microscopy/spectroscopy measurements reveal the very similar atomic structures and electronic properties, *e.g.* electronic states, and surface work functions, for all these intermetallic compound systems due to the physical and chemical similarities between these rare earth elements. Further, these electronic properties are periodically modulated by the moiré structures caused by the lattice mismatches between $ReAu_2$ and Au(111). These periodically modulated surfaces could serve as templates for the self-assembly of nanostructures. Besides, these two-dimensional rare earth-based intermetallic compounds provide platforms to investigate the rare earth related catalysis, magnetisms, etc., in the lower dimensions.

**TOC Graphic**

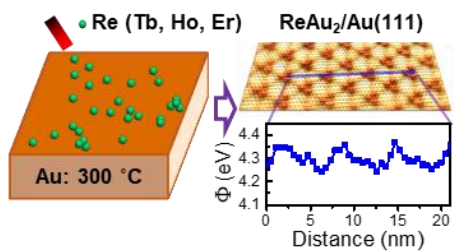

Surface alloying has attracted intense interest among physicists and material scientists because it is a straightforward route to control and modify the structure and electronic properties of surfaces,[1–3] which is fundamentally important for developing technologically relevant materials. Firstly, surface alloying provides an effective way to tune the chemical compositions in the surface layers and thus to tune chemical and physical properties of the relevant materials. For instance, for most catalysts, the active sites are located on or near the surface layers, and thus the catalytic performance could be tuned by varying the chemical composition in the surface layers.[4–7] Secondly, the adlayers formed by surface-confined alloying generally have different structures compared with the host materials. Such lattice mismatch results in moiré patterns in these systems, leading to periodically modulated surfaces that could be served as templates for self-assembled growth of nanostructures, *e.g.* nanocluster or molecules.[8] Further, the periodical modulations in the coupling between the adlayer and the subsequent layer, *e.g.* substrate, might give rise to new exotic phenomena.

Lanthanide rare earth-based intermetallic compounds on noble metal surfaces are good examples of periodically modulated surfaces by surface alloying. Deposition of gadolinium (Gd) on hot Au(111) or Ag(111) gives rise to the formation of long-range ordered $GdAu_2$ and $GdAg_2$.[9,10] The lattice mismatches in these systems results in periodically modulated moiré structures, leading to electronic modulations in these systems.[11,12] The structurally robust moiré lattices in $GdAu_2$ has been reported as templates for well-aligned organic nanowires,[6] and dense arrays of Co nanodots.[13–15] In addition, two-dimensional ferromagnetic orders have been reported in these systems mainly originated from the 4*f* electrons in the rare earth atoms.[10,16] The curie temperatures are ~10 K and ~80 K for $GdAu_2$ and $GdAg_2$, respectively. Further, the reduced dimensionality gives rise to new exotic states in these systems. For instance, Weyl nodal lines were observed in GdAg$_2$/Ag(111) by angular resolved photoemission spectroscopy (ARPES) combined with first-principle calculations.[17]

Besides Gd, other lanthanide element based intermetallic compounds by the same or similar approaches have also been reported, like LaAu$_2$,[18] CeAu$_2$,[18] and ErCu$_2$.[19] In this work, we apply the high-temperature growth procedure to another three lanthanide elements – terbium (Tb), holmium (Ho), and erbium (Er), on Au(111). These systems are particularly attractive because these rare earth elements have similar chemical bonds but have different 4$f$ electrons compared with La, Ce, and Gd, thus leading to different magnetisms. Further, these elements have different nuclear spins and nuclear magnetic moments, which might give rise to new magnetic phenomena due to the coupling between the nuclear and electron spins. Thus, these two-dimensional systems provide platforms to investigate rare earth related magnetisms. Via scanning tunneling microscopy (STM), we unveil that all these rare earth elements form very similar structures in the forms of TbAu$_2$, HoAu$_2$, and ErAu$_2$, similar to the reported rare earth -based intermetallic compound on Au(111). In addition, scanning tunneling spectroscopy (STS) measurement reveals very similar electronic properties in these intermetallic compounds, including local density of states and the surface work function. Further, the electronic states as well as the surface work function are periodically modulated by the moiré structures originating from the lattice mismatch.

Deposition of rare-earth metals onto clear Au(111) surfaces at 300 ˚C leads to the formation of well-ordered rare-earth gold surface alloys. **Figure 1a** presents a large-scale topographic image of TbAu$_2$ formed on Au(111) (for details on the determination of the stoichiometry, see **Section 1** in the supporting information), showing a highly ordered hexagonal superstructure, namely moiré structure, with lattice constants of 3.68 ± 0.10 nm. The inset in **Figure 1a** clearly shows three different regions within each unit cell of the moiré structure, dark spot, brightest spot, and the second brightest spot (referred as valley, hill, and bridge regions respectively in the following). The apparent height differences are revealed by the height profile across the unit cells, as shown in **Figure 1b**. The peaks and valleys in the height profile correspond to the hill and valley regions, whereas the shoulders adjacent to the peaks correspond to bridge region. The overall height differences vary with the sample bias (see **Figure S2** in the supporting information) from ~60 pm to ~40 pm, revealing these height differences are contributed from the geometric corrugation as well as electronic modulation, where the former one is dominant. Besides the moiré structure, the atomically resolved STM image (**Figure 1c**) shows a honeycomb structure with a lattice constant of 5.4 ± 0.2 Å, a factor of 4√3 smaller than that of moiré structure within the error limit. The unit cell is sketched out in the zoomed-in image (inset in **Figure 1c**), showing that only Au atoms were resolved in the atomic-resolution image. It should be noted that only Tb atoms could be resolved under some tunneling conditions. The corresponding fast Fourier transform (FFT) image present in **Figure 1d** shows the patterns for these two hexagonal structures, where the green and black vectors indicate the reciprocal lattices for the atomic lattice and moiré lattice. The angle between these two vectors was measured to be 30.0 ± 0.2°. Thus, the moiré structure is 4√3×4√3-R30° respective to the TbAu$_2$ lattice.

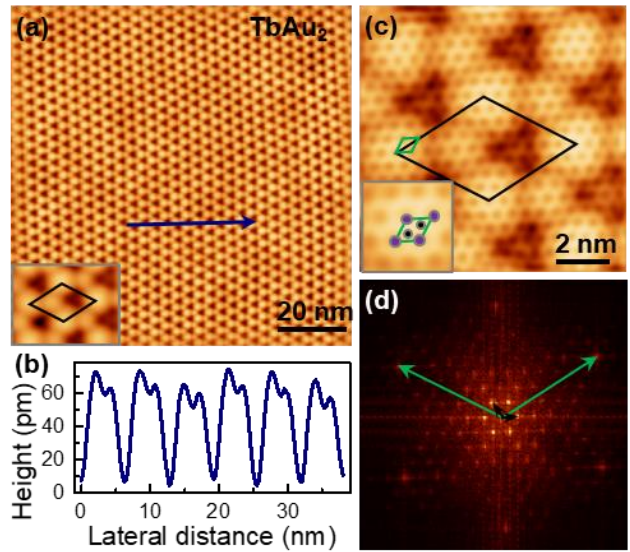

**Figure 1.** STM topographic images of TbAu$_2$. (a) Large-scale and (c) atomically resolved STM images of TbAu$_2$ on Au(111), respectively. (b) Height profile along the blue arrow in (a). (d) FFT of atomically resolved STM image of TbAu$_2$, the same area as (c) but with dimensions of 30 × 30 nm$^2$. Inset in (a) shows its zoom-in image, and the black diamond in the inset denotes the unit cell. Green and black diamonds in (c) denote the unit cell of the TbAu$_2$ and the moiré lattices. Inset in (c) shows its zoom-in image, where the purple and black solid circles represent Tb and Au atoms respectively. Green and black arrowed lines in (d) denote the reciprocal lattice vectors of the TbAu$_2$ and the moiré structures. Tunneling conditions for (a) U = 2.0 V, I = 200 pA, and (c) U = 0.5 V, I = 2.0 nA.

To identify the relation between the TbAu$_2$ and the Au substrate, we carried out the low energy electron diffraction (LEED) measurement. For the clean Au(111), the pattern shows only six spots as shown in **Figure 2a**, indicating the 3-fold symmetry as well as the cleanness of the Au substrate. In contrary, **Figure 2b** shows a set of moiré pattern for TbAu$_2$/Au(111). Compared with the pattern for clean Au(111), it reveals the angle of 30° between lattices of TbAu$_2$ and the moiré structure, in consistence with the atomic resolution STM results. Besides, the lattice vectors of the moiré structure are parallel to those of Au(111), and hence the angle between the lattices of TbAu$_2$ and Au(111) is 30°. Therefore, combined with the STM results, the moiré structure could be modeled as 4√3×4√3-R30° TbAu$_2$ on 13×13 Au(111).

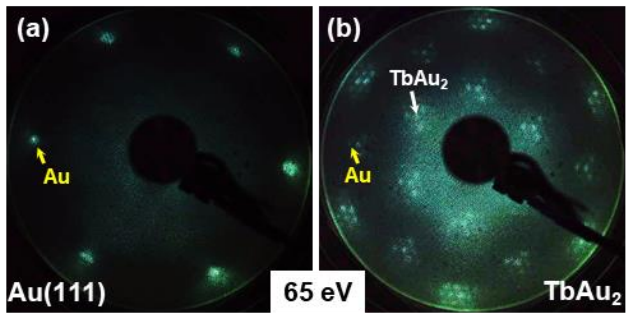

**Figure 2.** LEED patterns of (a) clean Au(111) and (b) TbAu$_2$/Au(111).

Similar structures have been reported in other rare-earth metal surface alloys, like $GdAu_2/Au(111)$,[20] $GdAg_2/Ag(111)$,[10] $LaAu_2/Au(111)$,[18] $CeAu_2/Au(111)$,[18] and $ErCu_2/Cu(111)$,[19] and so on. Naturally, the question comes out - whether other lanthanide metals have similar surface alloys on noble metal surfaces. Therefore, we employed the same approach for another two lanthanide elements – Ho and Er. **Figure 3** presents the STM results for the as grown $HoAu_2$ formed on Au(111) by deposition of Ho onto the Au(111) at 300 ˚C. It shows a highly ordered hexagonal moiré structure (**Figure 3a**) with a periodicity of $3.66 \pm 0.05$ nm. The atomic-resolution STM image in **Figure 3b** reveals the atomic lattice superimposed with the moiré structure. The unit cell is sketched out in the zoomed-in image (inset in **Figure3b**), showing only Ho atoms are resolved. The lattice constant for $HoAu_2$ was measured to be $5.4 \pm 0.1$ Å, a factor of $4\sqrt{3}$ smaller than that of moiré structure within the error limit. The corresponding FFT image present in **Figure 3c** shows the patterns for these two hexagonal structures. The unit cells in **Figure 3b** and the reciprocal lattice vectors in **Figure 3c** imply that the moiré lattice is $4\sqrt{3}\times4\sqrt{3}$-R30˚ respective to the atomic lattice of $HoAu_2$.

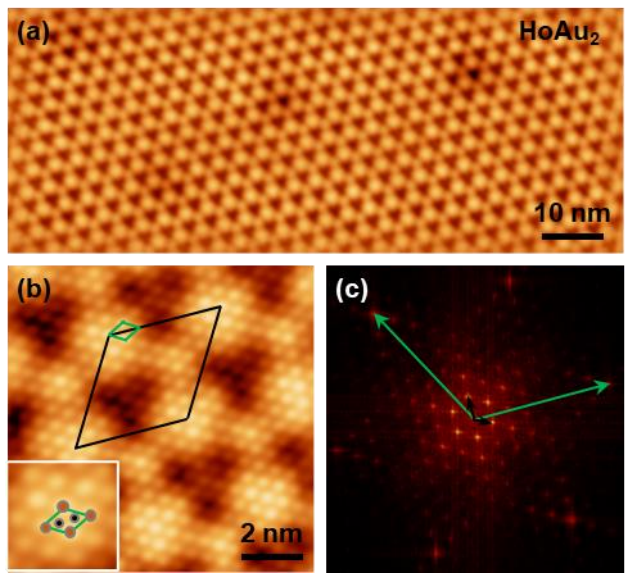

**Figure 3.** STM topographic images of $HoAu_2$. (a) Large-scale and (b) atomically resolved STM images of $HoAu_2$ on Au(111), respectively. Inset in (b) shows its zoom-in image, where the orange

and black solid circles represent Ho and Au atoms respectively. Green and black diamonds in (b) represent the unit cells of HoAu$_2$ and the moiré structures, respective. (c) FFT image of atomically resolved STM image of HoAu$_2$, the same area as (c) but with dimensions of 40 × 40 nm$^2$. Green and black arrowed lines in (c) denote the reciprocal lattice vectors of the HoAu$_2$ and the moiré structures. Tunneling conditions for (a) U = 2.0 V, I = 500 pA, and (b) U = 0.1 V, I = 500 pA.

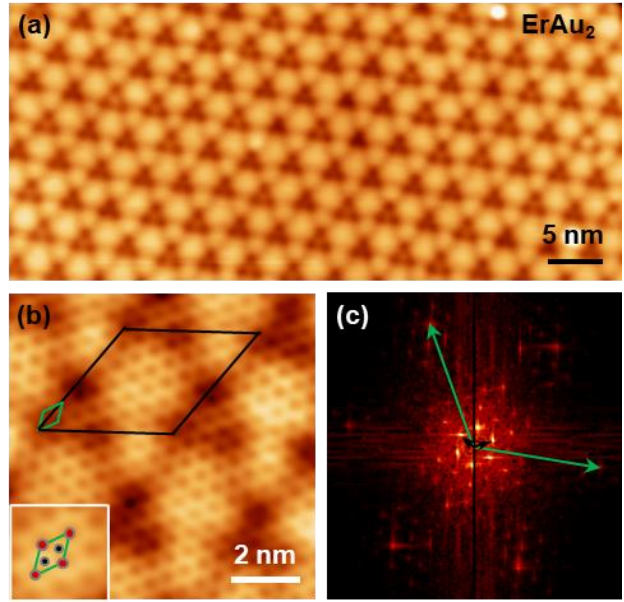

**Figure 4.** STM topographic images of ErAu$_2$. (a) Large-scale and (b) atomically resolved STM images of ErAu$_2$ on Au(111), respectively. Inset in (b) shows its zoom-in image, where the red and black solid circles represent Er and Au atoms respectively. Green and black diamonds in (b) represent the unit cells of ErAu$_2$ and the moiré structures, respective. (c) FFT image of atomically resolved STM image of ErAu$_2$, the same area as (c) but with dimensions of 25 × 25 nm$^2$. Green and black arrowed lines in (c) denote the reciprocal lattice vectors of the ErAu$_2$ and the moiré structures. Tunneling conditions for (a) U = -2.0 V, I = 200 pA, and (b) U = 5 mV, I = 2.5 nA.

Similar results were observed for ErAu$_2$ on Au(111) formed by deposition of Er atoms onto hot (300 °C) Au(111) substrate, as shown in **Figure 4**. The large-scale STM image (**Figure 4a**) reveal the long-range ordered moiré structures of ErAu$_2$/Au(111) with periodicity of 3.76 ± 0.04 nm. The atomic-resolution STM image in **Figure 4b** reveals the atomic lattice superimposed with the moiré structure. The unit cell is sketched out in the zoomed-in image (inset in **Figure4b**), showing only the Au atoms are resolved. The lattice constant for ErAu$_2$ was measured to be 5.4 ± 0.1 Å, a factor of 4√3 smaller than that of moiré structure within the error limit. The corresponding FFT image present in **Figure 4c** shows the patterns for these two hexagonal structures. The unit cells in **Figure 4b** and the reciprocal lattice vectors in **Figure 4c** imply that the moiré lattice is 4√3×4√3-R30° respective to the atomic lattice of ErAu$_2$. Therefore, TbAu$_2$, HoAu$_2$, and ErAu$_2$ on Au(111) have almost the same structures including the atomic lattices and moiré lattices, which is in consistent with the physical and chemical similarities between Tb, Ho, and Er elements.

For better comparison, we listed the structures for all the reported rare earth-based intermetallic compounds on Au(111) formed by surface alloying in **Table 1**. A hexagonal surface alloy could be formed for all the listed lanthanum elements, which is in consistent with the similarities of the lanthanum elements. The lattice constant for these hexagonal surface alloys slightly varies from 5.3 Å to 5.5 Å due to the slight differences in the atom size for the lanthanide elements. Accordingly, the lattice constant for the moiré structure varies from 3.2 nm to 3.8 nm due to the difference in the lattice mismatch between the surface alloy and Au(111) substrate. The moiré structure with different periodicities might be used to tune the growth of nanoclusters or self-assembles of molecules.

**Table 1.** Structures for rare earth-based intermetallic compounds on Au(111).

| REAu$_2$ | Structure | Atomic lattice (Å) | Moiré lattice (nm) | References |
|---|---|---|---|---|
| LaAu$_2$ | hexagonal | 5.3 | 3.2 | Ref. 18 |
| CeAu$_2$ | hexagonal | 5.4 | 3.3 | |
| GdAu$_2$ | hexagonal | 5.4~5.5 | 3.6~3.8 | Ref. 12, 16, 20 |
| TbAu$_2$ | hexagonal | 5.4 | 3.68 | |
| HoAu$_2$ | hexagonal | 5.4 | 3.66 | This work |
| ErAu$_2$ | hexagonal | 5.4 | 3.76 | |

So far, we have investigated the structures of the rare-earth gold surface alloys formed on Au(111) including TbAu$_2$, HoAu$_2$, and ErAu$_2$. They have almost the same hexagonal structures with the same lattice constant, and thus giving rise to the moiré structures with the same periodicity within the error limit. In the following, we will employ STS measurement to unravel the electronic properties of the rare-earth gold surface alloys, mainly the TbAu$_2$ on Au(111), modulated by the moiré structures.

**Figure 5a** presents the differential conductance spectra of TbAu$_2$ taken at three different regions within the unit cell of the moiré structure. All the spectra show similar features with a pronounced peak around 0.6 eV above the Fermi level. Similar features are observed in the other rare-earth metal surface alloys including HoAu$_2$/Au(111) and ErAu$_2$/Au(111) (see **Figures S3** and **S4** in the supporting information) and ErCu$_2$/Cu(111). The STS spectra for these systems show a pronounced peak around 0.6 eV, implying that this peak is mainly contributed from the rare-earth atoms in the alloys, slightly depending on the type of lanthanide atoms. Our theoretic calculations reveal that such state is mainly contributed from the d states of Tb atoms for the TbAu$_2$/Au(111) (see **Figure S5** in the supporting information). In the ErCu$_2$,[19] the peak around 0.6 eV is contributed from the Er d state mixed with Cu p state and d state confirmed by first-principle calculations. In addition, the energy for this peak is position-dependent due to the moiré modulations. In contrast, for TbAu$_2$ in this work, no energy shift was observed for the pronounced peak at ~0.6 eV at different regions, which might be due to the different coupling strength with the substrate. Instead, the intensity of this peak varies within the unit cell of the moiré structures. It shows higher intensity of this peak for the spectrum taken at the dark regions or the valley sites than the other two regions, indicating the higher local density of states (LDOS) at the valley regions within the unit cell of the moiré structures. To further confirm such position-dependence of the LDOS, we mapped the STS differential conductance over the unit cells of the moiré structures. As shown in **Figure 5b**, the dI/dV mapping at the energy of 0.6 eV clearly unveils the higher LDOS at the valley regions within the unit cell of the moiré structures, indicating the electronic state at 0.6 eV is modulated by the moiré structures. Besides, the spectra show overall slightly higher intensity at the valley regions below the energy level of ~1.2 eV than that in the hill and bridge regions, whereas slightly lower intensity above ~1.2 eV. The dI/dV mappings at -1.5 eV (**Figure 5c**) and 1.5 eV (**Figure 5d**) reveal the reversed contrast for the valley regions, illustrating the relative change in the LDOS due to the moiré modulations.

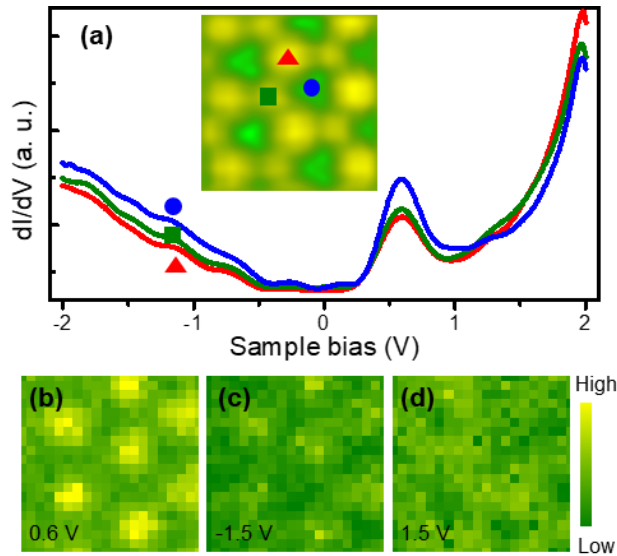

**Figure 5.** STS differential conductance spectra and mappings of TbAu$_2$/Au(111). (a) dI/dV spectra of TbAu$_2$ taken at three different regions (indicated in the inset STM image with area of $9 \times 9$ nm$^2$) within the unit cell of the moiré structure of TbAu$_2$. (b-d) Differential conductance mappings of TbAu$_2$ at sample bias at 0.6 V, -1.5 V, and 1.5 V, respectively. The corresponding topographic

image is present as the inset in (a). The tunneling conditions for the spectra and mappings are U = 2.0 V, I = 1.0 nA.

Further, we employed the STS in the distance-voltage z(V) mode to investigate the electronic structures of the TbAu$_2$/Au(111), which has been employed to study the field-emission resonance (FER) phenomenon.[21,22] FER provides an ideal playground for the study of fundamental physical properties like surface work function.[23] **Figure 6a** presents the dz/dV spectra of TbAu$_2$/Au(111) and clean Au(111) for comparison. On clean Au(111), it shows a pronounced step around 3.5 eV, corresponding the upper edge of the Au L-gap.[24,25] Besides, it reveals 7 FER states within the energy range to 10 eV. The first FER state locates at ~5.2 eV, which could be used as the reference for the surface work function. The more accurate surface work function can be calculated by the following equation based on a 1D tunneling junction model:[26]

$$E_n = \Phi + \left(\frac{3\pi\hbar}{2\sqrt{2m}}F\right)^{2/3} n^{2/3} \qquad (1)$$

Where $E_n$ is the energy of the $n^{th}$ FER state, $\Phi$ is the local surface work function, $\hbar$ is the reduced Planck constant, $m$ is effective mass of surface charge, and $F$ is the electric field. The surface work function of Au(111) was calculated to be 5.49 ± 0.09 eV by fitting the FER states according the equation (1) (**Figure 6b**), which agrees well with the reported value (5.35 eV).[27]

On TbAu$_2$/Au(111), all the spectra show 9 FER states within the scanning range of 10 eV with a weak dependence with position within the unit cell of the moiré structures. The first FER state locates around 2.6 eV (vs 5.2 eV for Au), indicating a reduction in the surface work function compared with clean Au(111). By fitting the FER states according the equation (1), as shown in **Figure 6b**, the surface work functions on TbAu$_2$/Au(111) were extracted to be 4.33 ± 0.11 eV, 4.28 ± 0.11 eV, and 4.25 ± 0.12 eV for the hill, bridge, and valley regions respectively. These values are in between the work functions of pure Tb (~ 3.1 eV) and gold (~5.4 eV), which is rational according to the composition dependence of work function in binary metal alloys.[28] Moreover, similar reduction in work function has been observed in the GdAu$_2$/Au(111) system by photoemission spectroscopy measurement.[6]

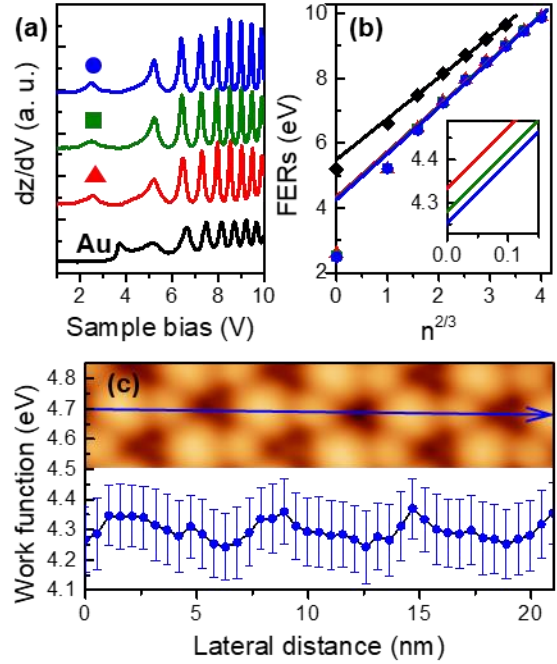

**Figure 6.** Surface work function of TbAu$_2$/Au(111). (a) dz/dV spectra of Au(111) and TbAu$_2$ taken at the featured regions within the unit cell of the moiré structures. The solid triangle, square, and circle indicate the positions where the spectra were taken, are the same as those in **Figure 3a**. (b) relations between the energy and index of the FER states. The energies of the FER states were extracted from the spectra in (a). The solid lines are the linearly fitted curves for n ≥ 2. The intersects at n = 0 are zoomed in and shown in the inset. (c) Surface work function profile of TbAu$_2$ along the blue arrow in the top panel. All the spectra were taken at the tunneling current of 20 pA to reduce the stark shift in the spectra.

To confirm the slight differences in the surface work functions of TbAu$_2$/Au(111) on different regions are caused by the modulation of the moiré structures rather than measurement uncertainty, we carried out the dz/dV line-mapping crossing several unit cells of the moiré structures. **Figure 6c** presents the surface work function profile along the blue arrow crossing three unit cells of the moiré structures of $TbAu_2$/Au(111). The surface work function of $TbAu_2$/Au(111) periodically varies with the same period of the moiré structures, revealing that the surface work function of $TbAu_2$/Au(111) is modulated by the moiré structures in consistence with the previous dI/dV results. It is worth to note that the surface work function of $HoAu_2$/Au(111) and $ErAu_2$/Au(111) was measured to be around 4.3 eV and modulated by the moiré structures too (see **Figures S7** and **S8** in the supporting information).

In summary, we have investigated the structural and electronic properties of three novel rare earth-based intermetallic compounds, namely $TbAu_2$, $HoAu_2$, $ErAu_2$, on Au(111) via STM/STS. The deposition of rare-earth metals onto the hot Au(111) surface gives rise to the formation of long-range ordered monolayers of intermetallic compounds. Atomically resolved STM images reveal hexagonal lattices with lattice constant of ~5.4 Å for all these monolayers on Au(111), superimposed with moiré structures. The lattice constant for the moiré structures slightly depends on the type of lanthanide atoms, in the range of 3.66~3.76 nm. The moiré structures could be modeled as $4\sqrt{3}\times4\sqrt{3}$-R30° $ReAu_2$ (Re = Tb, Ho, and Er) on 13×13 Au(111). STS measurement illustrated the very similar electronic structures for all these intermetallic compound systems with a pronounced state at ~0.6 eV above the Fermi level. Besides, the FER spectra reveal a significant reduction in the surface work function after the formation of intermetallic compounds. Further, the electronic states and surface work function are periodically modulated by the moiré structures. These periodically modulated surfaces could serve as templates for the self-assembly of organic molecules or nanoclusters. In addition, these two-dimensional rare earth-based intermetallic compounds provide platforms to investigate the rare earth related catalysis, magnetisms, etc., in the lower dimensions.

**MATERIALS AND METHODS**

*Sample preparation.* Clean Au(111) surfaces were prepared by repeated cycles of $Ar^+$ sputtering (600 eV) and annealing at 450 °C. The cleanness of the Au(111) surface was verified by STM topographic images. Rare-earth metals (Tb, Ho, and Er) were deposited onto clean Au(111) via a commercial e-beam evaporator (EFM-3T, omicron). During the deposition, the Au(111) substrate was kept at 300 °C to stimulate the formation of well-ordered rare-earth-gold surface alloys.

All the experiments were carried out in an UHV LT-STM system with a base pressure better than $2 \times 10^{-10}$ mbar. The sample was directly transferred to the STM without breaking the UHV environment after the growth of the surface alloys. For the LEED measurement, samples were transferred to a separated UHV VT-STM system equipped with LEED setup using a home-made UHV suitcase. The cleanness of the sample after the transfer was verified by STM imaging.

*LEED and STM/STS measurement.* LEED measurements were conducted at room temperature with an incident electron energy of 65 eV. The STM images were acquired in a constant-current mode at the temperature of liquid nitrogen (∼78 K) using an electrochemically etched tungsten tip and all the labelled bias voltages referred to the sample against tip. All the STM images were post-processed by the WSxM software.[29] The dI/dV (dz/dV) spectra and mapping were obtained by numerical derivation of I(V) [Z(V)] spectra. Prior to the spectroscopic measurement, the STM tip was calibrated against the surface states of the Au(111) or Cu(111) surfaces.

## ASSOCIATED CONTENT

**Supporting Information**

Details of Determination of stoichiometry of the rare earth – gold intermetallic compounds, Bias-dependent height corrugations in TbAu$_2$/Au(111), Electronic properties of HoAu$_2$/Au(111) and ErAu$_2$/Au(111), Band structures of TbAu$_2$, Stark shift in the FERs of TbAu$_2$/Au(111), Surface work function of HoAu$_2$/Au(111) and ErAu$_2$/Au(111).

## AUTHOR INFORMATION

**Corresponding Authors**

*Email: ydque@phy.cuhk.edu.hk and xdxiao@phy.cuhk.edu.hk

**ORCID**

Yande Que: 0000-0002-5267-4985

Chaoqiang Xu: 0000-0001-8561-8974

Xudong Xiao: 0000-0003-0551-1144

**Notes**

The authors declare no competing financial interest.

## ACKNOWLEDGEMENTS

This work was supported by the Research Grant Council of Hong Kong (No. 404613), the Direct Grant for Research of CUHK (No. 4053306 and No. 4053348), the National Natural Science Foundation of China (No. 11574128) and the MOST 973 Program (No. 2014CB921402).

# Supporting information

# Two-Dimensional Rare Earth - Gold Intermetallic Compounds on Au(111) by Surface Alloying

Yande Que,[1*] Yuan Zhuang,[1] Ziyuan Liu,[2] Chaoqiang, Xu,[1] Bin Liu,[1] Kedong Wang,[3] Shixuan Du,[2] and Xudong Xiao[1*]

[1] Department of Physics, the Chinese University of Hong Kong, Shatin, Hong Kong, China

[2] Institute of Physics and University of Chinese Academy of Sciences, Chinese Academy of Sciences, Beijing, 100190, China

[2] Department of Physics, Southern University of Science and Technology, Shenzhen, Guangdong 518055, China

[*] Corresponding author.

Email: ydque@phy.cuhk.edu.hk and xdxiao@phy.cuhk.edu.hk

**S1. Determination of stoichiometry of the rare earth – gold intermetallic compounds**

In the previous work,[1] the stoichiometry, namely, the ratio between the rare earth and gold atoms in the intermetallic compound alloys was determined by high-resolution X-ray photoemission electron spectroscopy (XPS), combined with atomically resolved STM imaging. In this work, atomic-resolution STM images (**Figure S1**) shows two different contrasts for the surface alloys, namely, honeycomb lattices and triangular lattices, corresponding to the gold atoms and rare earth atoms in the surface alloy, respectively. Such results have also been observed in other rare earth-based intermetallic compound alloys, like $GdAu_2$,[1] $ErCu_2$,[2] etc. Although STM is not a chemical sensitive technique, considering the physical and chemical similarities between the rare earth elements, still, it is reasonable to associate the observed structures for the surface alloy in **Figure S1** to $TbAu_2$, despite lack of further direct evidences.

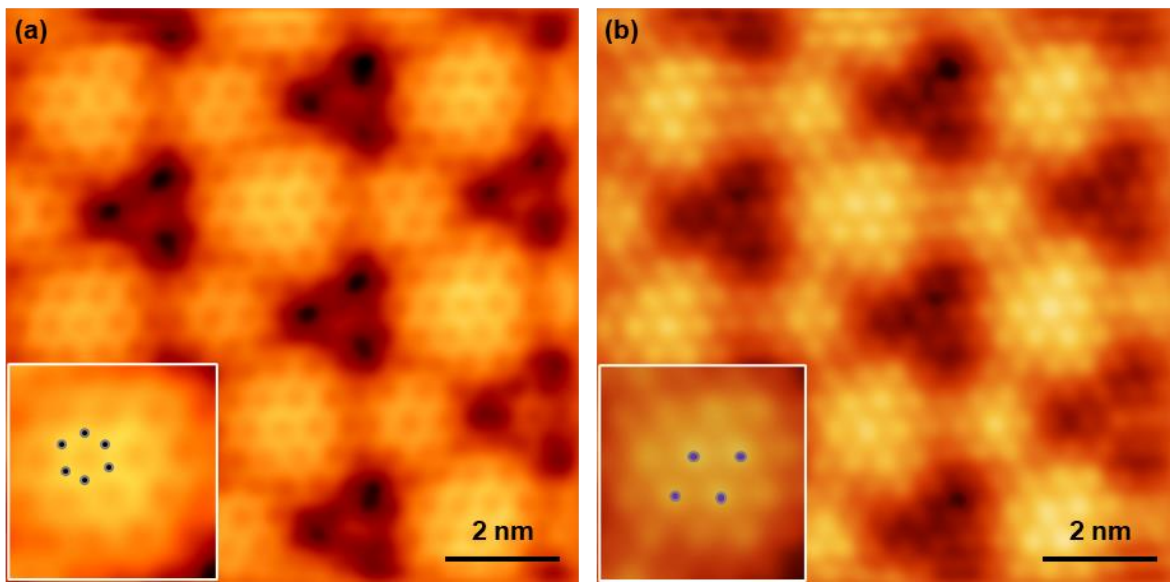

**Figure S1.** Atomic-resolution STM images of $TbAu_2$ monolayer on Au(111) under (a) 0.5 V and (b) -0.5 V, respectively. The insets show the zoomed-in images, showing honeycomb and triangular lattices, respectively. The tunneling current was set to be 2.0 nA for both (a) and (b).

**S2. Bias-dependent height corrugations in $TbAu_2$/Au(111)**

Generally, the height corrugations in the topographic STM images are contributed from the geometric corrugations and variations in the local density of electronic states (LDOS). The geometric corrugations only depend on the sample surface, and thus independent on the sample bias. In contrary, the LDOS is generally energy-dependent, namely bias-dependent. **Figure S2a** shows the height corrugations of TbAu$_2$/Au(111) at series of sample bias from -3.0 V to 3.0 V. It clearly reveals the variations of the height corrugations at different sample bias, indicating the height corrugations are partially contributed from the electronic modulations. Besides, the height corrugations almost remain the same as increasing the tunneling current from 0.1 nA to 2.0 nA, as shown in **Figure S2b**.

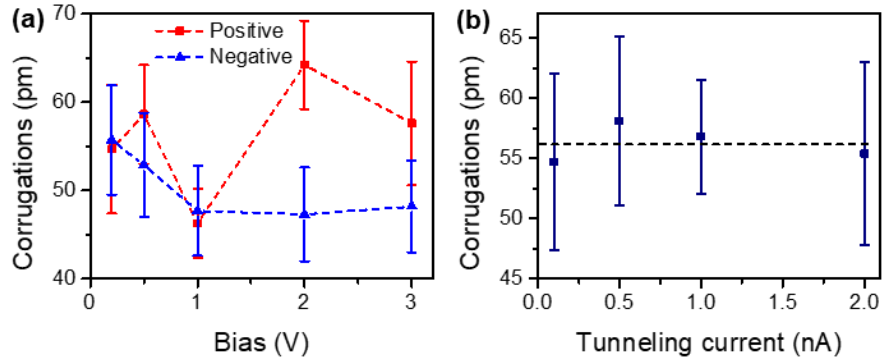

**Figure S2.** Bias-dependent height corrugations in TbAu$_2$/Au(111). Relationships between the height corrugations and (a) sample bias, and (b) tunneling current. Tunneling condition for (a): I$_{set}$ = 100 pA, (b) V = 0.2 V.

### S3. Electronic properties of HoAu$_2$/Au(111)

As shown in **Figure S3a**, All the spectra show a pronounced peak around 0.6 eV with higher intensity at the valley region than that at bridge and hill regions, similar as those for TbAu$_2$/Au(111) in the main text and ErAu$_2$/Au(111). Besides, spectra reveal overall higher intensity below ~1.3 eV at the valley regions compared that at the bridge and hill regions within the unit cell of moiré structures, indicating the modulations in LDOS due to the moiré structures. Such electronic modulations are further confirmed by the dI/dV mappings at 0.6 eV and ±1.5 eV, as shown in **Figures S3(c-e)**.

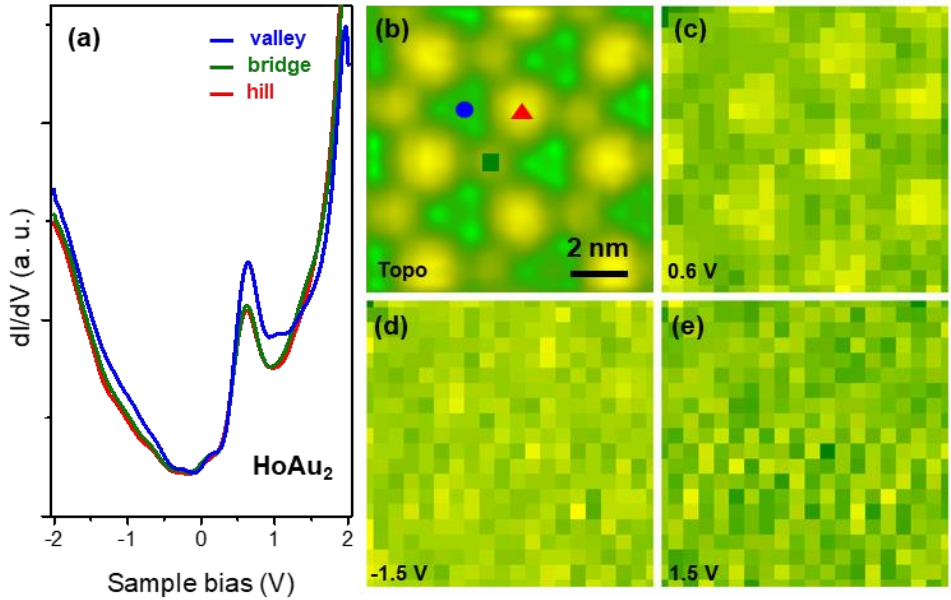

**Figure S3.** dI/dV spetra and mappings of HoAu$_2$/Au(111). (a) dI/dV spectra of HoAu$_2$ taken at three different regions (indicated in the STM image in (b) with area of $10 \times 10$ nm$^2$) within the unit cell of the moiré structure of HoAu$_2$. Topographic STM image (b) and differential conductance mapping of HoAu$_2$ at sample bias at 0.6 V (c), -1.5 V (d), and 1.5 V (d), respectively. The tunneling conditions for the spectra and mappings are U = 2.0 V, I = 1.0 nA.

## S4. Electronic properties of ErAu$_2$/Au(111)

As shown in **Figure S4a**, All the spectra show a pronounced peak around 0.6 eV with higher intensity at the valley region than that at bridge and hill regions, similar as those for TbAu$_2$/Au(111) in the main text. Besides, spectra reveal overall higher intensity below ~1.3 eV at the valley regions compared that at the bridge and hill regions within the unit cell of moiré structures, indicating the modulations in LDOS due to the moiré structures. Such electronic modulations are further confirmed by the dI/dV mappings at 0.6 eV and ±1.5 eV, as shown in **Figures S4(c-e)**.

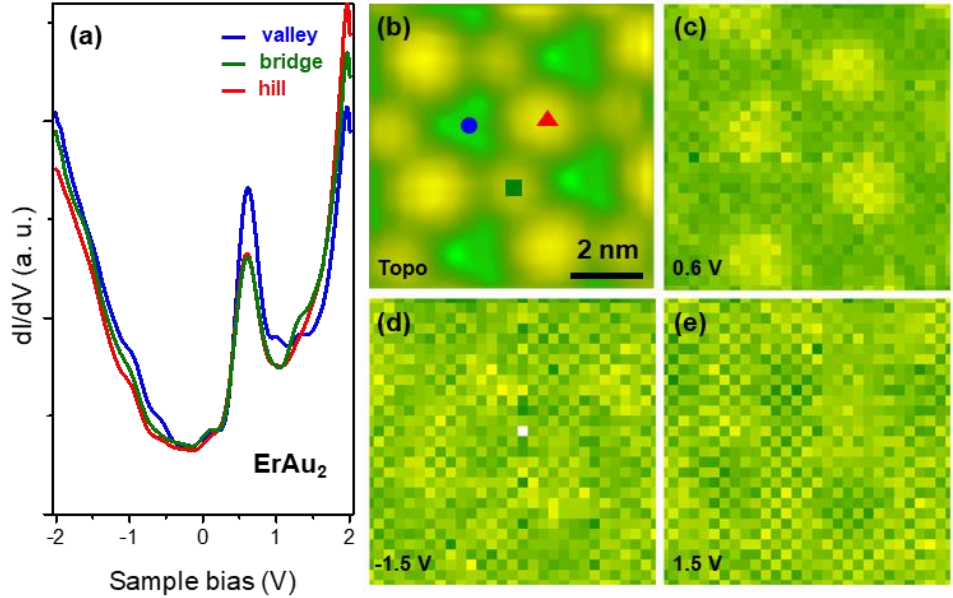

**Figure S4.** dI/dV spetra and mappings of ErAu$_2$/Au(111). (a) dI/dV spectra of ErAu$_2$ taken at three different regions (indicated in the STM image in (b) with area of $8 \times 8$ nm$^2$) within the unit cell of the moiré structure of ErAu$_2$. Topographic STM image (b) and differential conductance mapping of HoAu$_2$ at sample bias at 0.6 V (c), -1.5 V (d), and 1.5 V (d), respectively. The tunneling conditions for the spectra and mappings are U = 2.0 V, I = 1.0 nA.

## S5. Band structures of TbAu$_2$

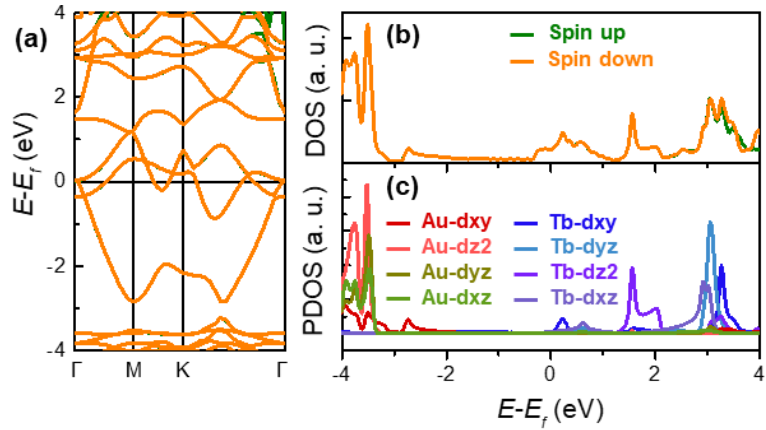

**Figure S5.** (a) Band structures of TbAu$_2$. The orange and green curves are bands for spin up and spin down, respectively. (b) Density of state (DOS) and (c) projected DOS (PDOS) of TbAu$_2$ within the range of ±4 eV.

The calculations on the structure and electronic states of TbAu$_2$ were based on density functional theory (DFT). We calculated the electronic structure using Vienna Ab-initio Simulation Package (VASP),[3,4] within the Projector augmented-wave (PAW)[5] method with plane wave basis and Perdew-Burke-Ernzerhopf (PBE) generalized-gradient approximation (GGA)[6] were used. The plane-wave cut-off energy of 650 eV and (9×9×1) k-points were used. The vacuum spacing of 20 Å along the z axis was used to avoid interlayer couplings. Because of the strongly correlated electrons in Tb 4f orbitals, we use GGA+U (U=6) to calculate the structure. The lattice constant of the TbAu$_2$ is 5.12 Å after the structural was optimized to the lowest energy, slightly smaller than the STM results which might be due to the absence of Au substrate. The band structures shown in **Figure S5a** reveal that the TbA$_2$ monolayer is metallic, namely zero-bandgap. Spin-resolved density of state (**Figure S5b**) shows no unpaired spins near the Fermi energy. Further magnetic measurement is required to reveal the magnetic properties of TbAu$_2$/Au(111) as well as another two rare earth-gold intermetallic compounds. In addition, it reveals a pronounced peak in the DOS around 1.6 eV for both spin up and spin down states, which is mainly contributed from the dz2 states of Tb atoms in the system (**Figure S5c**). Hence, we could associate the peaks at 0.6 eV in the dI/dV spectra in the main text to this peak, mainly contributed from the d states of Tb atoms. The difference the calculation results and experiment STS results might originate from the simplicity of the calculation model, *e.g.* absence of the substrate.

## S6. Stark shift in the FERs of TbAu$_2$/Au(111)

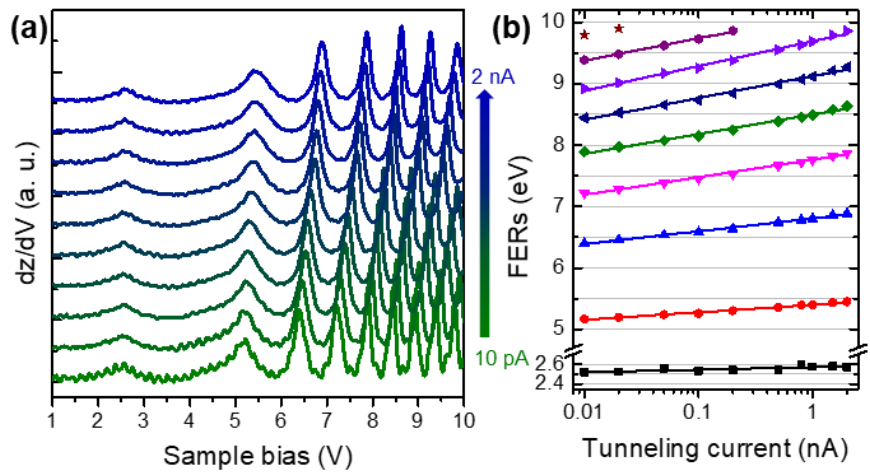

**Figure S6**. Stark shift in the FERs of TbAu$_2$/Au(111). (a) A series of dz/dV spectra of TbAu$_2$/Au(111) taken at the hill region under the tunneling current from 20 pA to 2.0 nA. The spectra are vertically shifted evenly for clarity. (b) Relations between the FERs and the tunneling current. The solid lines are the linear fit under the semi-log in the current axis.

Electric field applied by the STM tip will induce Stark shifts in the FERs of TbAu$_2$ on Au(111). **Figure S6a** present a series of dz/dV spetra under tunneling current from 10 pA to 2 nA. It clearly shows the shifts in the FERs as increasing the tunneling current. The energy positions of FERs are plotted in **Figure S6b**. The linear fittings illustrate the relationship between the FERs and the tunneling current, which could be described by $E \propto lnI$. It is worthy to note that the first FERs around 2.5 eV remains almost unchanged with the tunneling current.

## S7. Surface work function of HoAu$_2$/Au(111)

**Figure S7a** present the line mapping of dz/dV spectra on HoAu$_2$/Au(111) along the line shown in **Figure S7b**. It shows slight variations in the energies of the FERs periodically caused by the moiré structures. The surface work functions were extracted by fitting these FERs according the equation (1) in the main text and presented in **Figure S7b**. It reveals peaks of surface work function at the hill regions, indicating the surface work function is periodically modulated by the moiré structures.

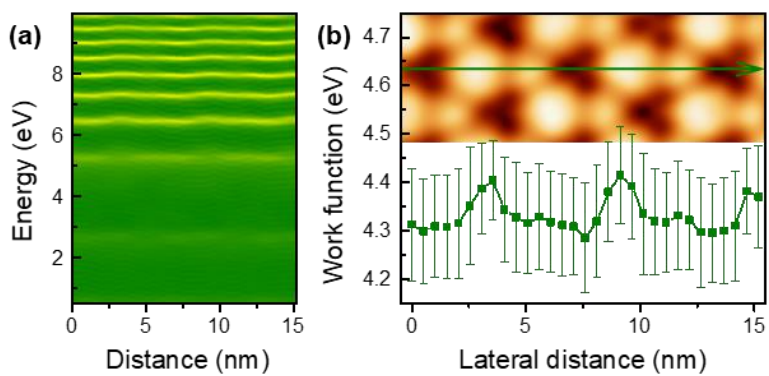

**Figure S7.** Surface work function of HoAu$_2$/Au(111). (a) Line mapping of dz/dV spectra and (b) Surface work function profile along the green line in top panel of (b). The surface work function was obtained by fitting the dz/dV spectra in (a) according to the equation (1) in the main text.

## S8. Surface work function of ErAu$_2$/Au(111)

**Figure S8a** present the line mapping of dz/dV spectra on ErAu$_2$/Au(111) along the line shown in **Figure S8b**. It shows slight variations in the energies of the FERs periodically caused by the moiré structures. The surface work functions were extracted by fitting these FERs according the equation (1) in the main text and presented in **Figure S8b**. It reveals peaks of surface work function at the hill regions, indicating the surface work function is periodically modulated by the moiré structures

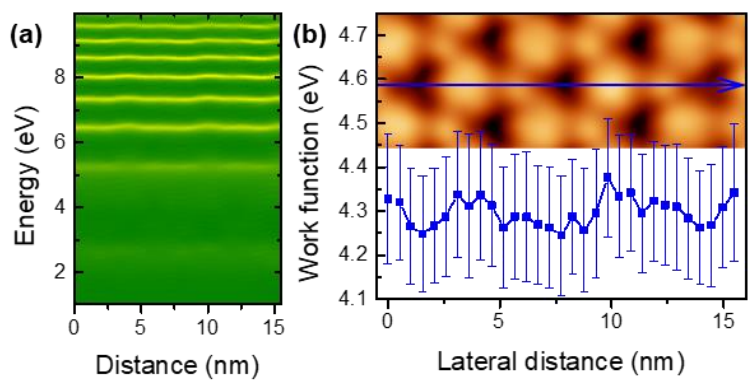

**Figure S8.** Surface work function of ErAu$_2$/Au(111). (a) Line mapping of dz/dV spectra and (b) Surface work function profile along the green line in top panel of (b). The surface work function was obtained by fitting the dz/dV spectra in (a) according to the equation (1) in the main text.